\documentclass[prl,twocolumn,floatfix,showpacs]{revtex4}
\usepackage[dvips]{graphicx}
\usepackage{amsmath}
\usepackage{url}
\usepackage{amssymb}
\usepackage{amsfonts}

\newcommand{\ion}[2]{\mbox{\ensuremath{^{#2}}#1\ensuremath{^+}}}

\newcommand{\Ca}[1]{\ion{Ca}{#1}}
\newcommand{\lev}[2]{\mbox{#1$_{\mbox{\tiny$#2$}}$}}
\newcommand{\hfslev}[3]{\mbox{\ensuremath{\textrm{#1}^{\mbox{\tiny\textrm{#3}}}_{\mbox{\tiny\textrm{#2}}}}}}

\newcommand{\ket}[1]{\ensuremath{\left| #1 \right>}}

\newcommand{\ketdn}{\ket{\downarrow}}
\newcommand{\ketup}{\ket{\uparrow}}

\newcommand{\unit}[1]{\,\mbox{#1}}
\newcommand{\Hz}{\unit{Hz}}
\newcommand{\kHz}{\unit{kHz}}
\newcommand{\GHz}{\unit{GHz}}
\newcommand{\mW}{\unit{mW}}
\newcommand{\s}{\unit{s}}
\newcommand{\us}{\unit{$\mu$s}}
\newcommand{\G}{\unit{G}}
\newcommand{\MpG}{\unit{MHz\,G$^{-1}$}}

\newcommand{\phierr}{\ensuremath{\phi_{\textrm{err}}}}
\newcommand{\id}{\,\mathrm{d}}  % This is a roman d for use in e.g. dx at the end of integrals.  It only works in math mode and includes a small space.
\newcommand{\myetal}{\textit{et al.}}
\newcommand{\pipulse}{\ensuremath{\pi}-pulse}  % Just an abbreviation really, but I may want to add a hyphen.
\newcommand{\piotwopulse}{\ensuremath{\frac{\pi}{2}}-pulse}  % Just an abbreviation really, but I may want to add a hyphen.
\newcommand{\pipulses}{\ensuremath{\pi}-pulses}  % Just an abbreviation really, but I may want to add a hyphen.
\newcommand{\piotwopulses}{\ensuremath{\frac{\pi}{2}}-pulses}  % Just an abbreviation really, but I may want to add a hyphen.
\newcommand{\eqn}[1]{(\ref{eqn:#1})}
\newcommand{\rme}{\mathrm{e}}

\newcommand{\rmd}{\mathrm{d}}

\newcommand{\degree}{\mbox{$^{\circ}$}}

\begin{document}

\title{Keeping a single qubit alive by experimental dynamic decoupling}

\author{D J Szwer\footnote{Present address: Department of Physics, Durham University, South Road, Durham DH1 3LE, UK.}, S C Webster, A M Steane and D M Lucas}
\affiliation{Department of Physics, University of Oxford, Clarendon Laboratory, Parks Road, Oxford OX1 3PU, UK.}
%\ead{david.szwer@gmail.com}

%\date{30 July 2010}
%\maketitle

\begin{abstract}
We demonstrate the use of dynamic decoupling techniques to extend the coherence time of a single memory qubit by nearly two orders of magnitude. By extending the Hahn spin-echo technique to correct for unknown, arbitrary polynomial variations in the qubit precession frequency, we show analytically that the required sequence of $\pi$-pulses is identical to the Uhrig dynamic decoupling (UDD) sequence. We compare UDD and CPMG sequences applied to a single \Ca{43} trapped-ion qubit and find that they afford comparable protection in our ambient noise environment.   
\end{abstract}


%\submitto{\jpb}

\maketitle

% Introduction

\begin{figure}[b!]
\centerline{\includegraphics[width=0.5\textwidth]{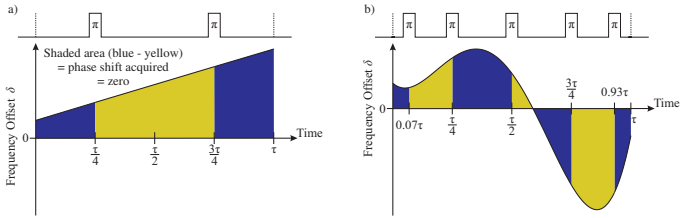}}
\caption{Illustration of the effect of dynamic decoupling sequences on the acquired phase, as calculated by \eqn{uhrigInt}.  a): Frequency offset $\delta$ varies linearly with time, and the phase shift can be completely corrected by a sequence with two \pipulses\ (for $n=2$, CPMG and UDD are identical).  b): $\delta$ varies as a (arbitrary, unknown) quartic polynomial, and is perfectly corrected by $n=5$ pulse UDD.  \label{fig:TheoryGraph}}
\end{figure}

Dynamic decoupling (DD) is a general technique for maintaining the phase coherence of a quantum state, with particular importance for protecting the quantum information stored in the memory qubits of a quantum computer \cite{Viola1998}.  The simplest example is the Hahn spin-echo \cite{Hahn1950}, a single \pipulse\ which protects against an arbitrary and unknown constant offset in the qubit's precession frequency \cite{Chiaverini2005, Lucas2007}.  When the state is subject to a time-varying offset due to, for example, magnetic field noise, it can be protected by a sequence of many \pipulses.  One of these, the Carr-Purcell-Meiboom-Gill (CPMG) sequence, is well known in the field of nuclear magnetic resonance \cite{Freeman1998}.  More recently, other sequences have been investigated specifically for their dynamic decoupling properties, such as Periodic DD, Concatenated DD \cite{Khodjasteh2005}, random decoupling \cite{Viola2005}, composite schemes \cite{Kern2005}, and local optimisation \cite{Biercuk2008, Uys2009}; a recent review by Yang, Wang and Liu contains further information and references \cite{Yang2010}.

In this paper, we derive a dynamic decoupling sequence in a particularly intuitive manner, as an extension to the spin-echo \cite{Hahn1950}.  We prove that with $n$ pulses, the sequence can cancel out all the dephasing that would be caused by the frequency varying as an $(n-1)$th order polynomial function of time, without knowledge of the polynomial coefficients.  This sequence is identical to the Uhrig Dynamic Decoupling (UDD) sequence \cite{Uhrig2007, Uhrig2008}, which was originally derived by considering the interaction of a spin qubit with a bosonic bath. We implement the sequence on a single \Ca{43} ion, demonstrating that the coherence time of this qubit is significantly increased, and compare it with the CPMG sequence.

Suppose an arbitrary qubit state is prepared at time 0, and we want to recover it at time $\tau$.  The pulse sequence is a series of (assumed ideal and instantaneous) \pipulses\ at times $\alpha_1\tau, \alpha_2\tau, \dotsc, \alpha_n\tau,$ where the $\alpha_i$ are to be found.  We have remarked that a single Hahn spin-echo will correct for a constant frequency offset.  If the offset varies linearly with time, we can correct the phase error with two \pipulses\ at $t=\frac{1}{4}$ and $\frac{3}{4}$, where $t=\textrm{time}/\tau$ (Figure \ref{fig:TheoryGraph}a).  To generalise further, postulate that $n$ pulses suffice to correct for a frequency variation $\delta(t)$ that is an $(n-1)$th-order polynomial in time (Figure \ref{fig:TheoryGraph}b):
\begin{equation}
	\delta(t) = p_0+p_1t+p_2t^2+\dotsb+p_{n-1}t^{n-1}.
\end{equation}
The phase error \phierr\ is given by integrating $\delta(t)$ over time.  But each \pipulse\ reverses the direction of the qubit's precession, so between pulses $i$ and $i+1$, if $i$ is odd, we multiply the acquired phase by $(-1)$.  The resulting integral is thus
\begin{eqnarray}
%	\begin{split}  % Can't use split from AMSmath...
		\phierr &= \sum^n_{i=0} (-1)^i \int^{\alpha_{i+1}}_{\alpha_i} \delta(t) \id t \nonumber\\
		  &= \sum^n_{i=0} (-1)^i \int^{\alpha_{i+1}}_{\alpha_i} \sum^n_{j=1} p_{j-1} t^{j-1} \id t \nonumber\\
			&= \sum^n_{i=0} (-1)^i \left[ \sum^n_{j=1} \frac{p_{j-1} t^j}{j} \right]^{\alpha_{i+1}}_{\alpha_i}
		\label{eqn:uhrigInt}
%	\end{split}
\end{eqnarray}
where $\alpha_0=0$ and $\alpha_{n+1}=1$.  Collecting terms for each polynomial coefficient $p_j$:
\begin{equation}
%	\begin{split}
		\phierr = \sum^n_{j=1} \frac{p_{j-1}}{j} \left[ (-1)^n - 2\sum^n_{i=1} (-1)^i \alpha_i^j \right].
%	\end{split}
\end{equation}
We require \phierr\ to be 0 for \emph{any} choice of the $p_j$, and so we obtain a set of $n$ simultaneous equations for the $\alpha_i$
\begin{equation}
	(-1)^n - 2\sum^n_{i=1} (-1)^i \alpha_i^j = 0 \quad \quad \forall j=1,2,\dotsc,n
\label{eqn:uhrigEqns}
\end{equation}
These are solved by
\begin{equation}
	\alpha_i=\sin^2\left(\frac{\pi}{2}\frac{i}{n+1}\right)
	\label{eqn:uhrigSin}
\end{equation}
which can be proved directly by substituting \eqn{uhrigSin} into \eqn{uhrigEqns} and applying a series of trigonometric identities \cite{Szwer2010}.

The sequence is independent of $\tau$; however in practice the frequency offset $\delta(t)$ is only \emph{approximated} by a polynomial, and as $\tau$ increases we need more polynomial terms (and hence more \pipulses) for the approximation to be valid.

This sequence was previously and independently discovered by Uhrig \cite{Uhrig2007, Uhrig2008}, by considering the spectral properties of a qubit coupled to a bath of bosons that cause decoherence.  The echo sequence was treated as a filter in frequency space.  Uhrig demanded that the first $n$ derivatives of the filter function vanish at zero frequency, because this gives the strongest suppression of the noise at low frequencies, and this condition leads to the simultaneous equations \eqn{uhrigEqns} and hence the sequence \eqn{uhrigSin}.  Lee, Witzel and Das Sarma have shown \cite{Lee2008} that this sequence is optimal for \emph{any} dephasing Hamiltonian, where ``optimal'' means that it is the sequence that maximises the qubit fidelity in the small $\tau$ limit, for a given number of pulses\footnote{We note that \eqn{uhrigSin} also gives the locations of the zeros of Chebyshev polynomials of the second kind $U_n(2t-1)$ (where the polynomials have been scaled and shifted from the domain $x\in[-1,1]$ to $t\in[0,1]$).}.  While this paper was in preparation, Hall \myetal\ have independently published a derivation equivalent to ours \cite{Hall2010}.

In a 1988 paper \cite{Keller1988}, Keller and Wehrli suggest using a theoretical procedure similar to ours, to cancel the effects of successive polynomial orders of fluid flow in MRI.  However, this ``gradient moment nulling'' allows $\delta(t)$ to be controllably scaled by the experimenter; Keller and Wehrli concentrate on this parameter rather than pulse timing and so do not find the UDD sequence.

% Experiment

The first experimental tests of UDD were by Biercuk \myetal, who applied a variety of dynamic decoupling schemes to ensembles of $\sim 1000$ \ion{Be}{9} ions in a Penning trap \cite{Biercuk2009}.  Dynamic decoupling was demonstrated in a solid by Du \myetal\ (using electron paramagnetic resonance of ensembles of unpaired carbon valence electrons in irradiated malonic acid crystals) \cite{Du2009}, and in a dense atomic gas by Sagi, Almog and Davidson ($\sim 10^6$ \mbox{\ensuremath{^{87}}Rb} atoms in a dipole trap) \cite{Sagi2010}.  And recently, Ryan, Hodges and Cory implemented sequences using single nitrogen vacancy centres in diamond \cite{Ryan2010}.

We have applied dynamic decoupling to a single \Ca{43} trapped-ion qubit, held in a radio-frequency Paul trap \cite{Barton2000}. The qubit is stored in two hyperfine states in the ground level, $\ketdn=\hfslev{4S}{1/2}{4,+4}$ and $\ketup=\hfslev{4S}{1/2}{3,+3}$ (where the superscripts indicate the quantum numbers $F,M_F$); these states are separated by a 3.2\GHz\ M1 transition.  The transition's sensitivity to the external magnetic field is 2.45\MpG\ at low field; we apply a field of 2.2\G\ to define a quantization axis and to increase the ion's fluorescence rate (by destabilising dark states \cite{Berkeland2002}). Rabi oscillations are driven on the qubit transition at Rabi frequency $2\pi\times18\kHz$, using microwaves. These are generated using a versatile synthesizer, amplified with a solid-state amplifier (to $\approx 750\mW$) and broadcast inside the vacuum chamber using a trap electrode as the antenna. To improve the fidelity of the dynamic decoupling \pipulses\ we apply a small 50\Hz\ signal, synchronized with the AC line, to a magnetic field coil which cancels the dominant component of the magnetic field fluctuations experienced by the ion; the remaining noise has amplitude up to $\pm 3\kHz$. Each experimental sequence is also line-triggered.
% 2.193(8)G in the linear Zeeman model.

\begin{figure}[b]
	\centerline{\includegraphics[width=0.5\textwidth]{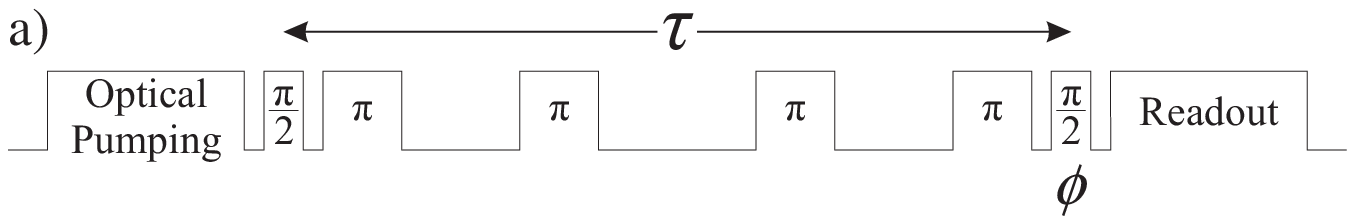}}
	\centerline{\includegraphics[width=0.35\textwidth, angle=-90]{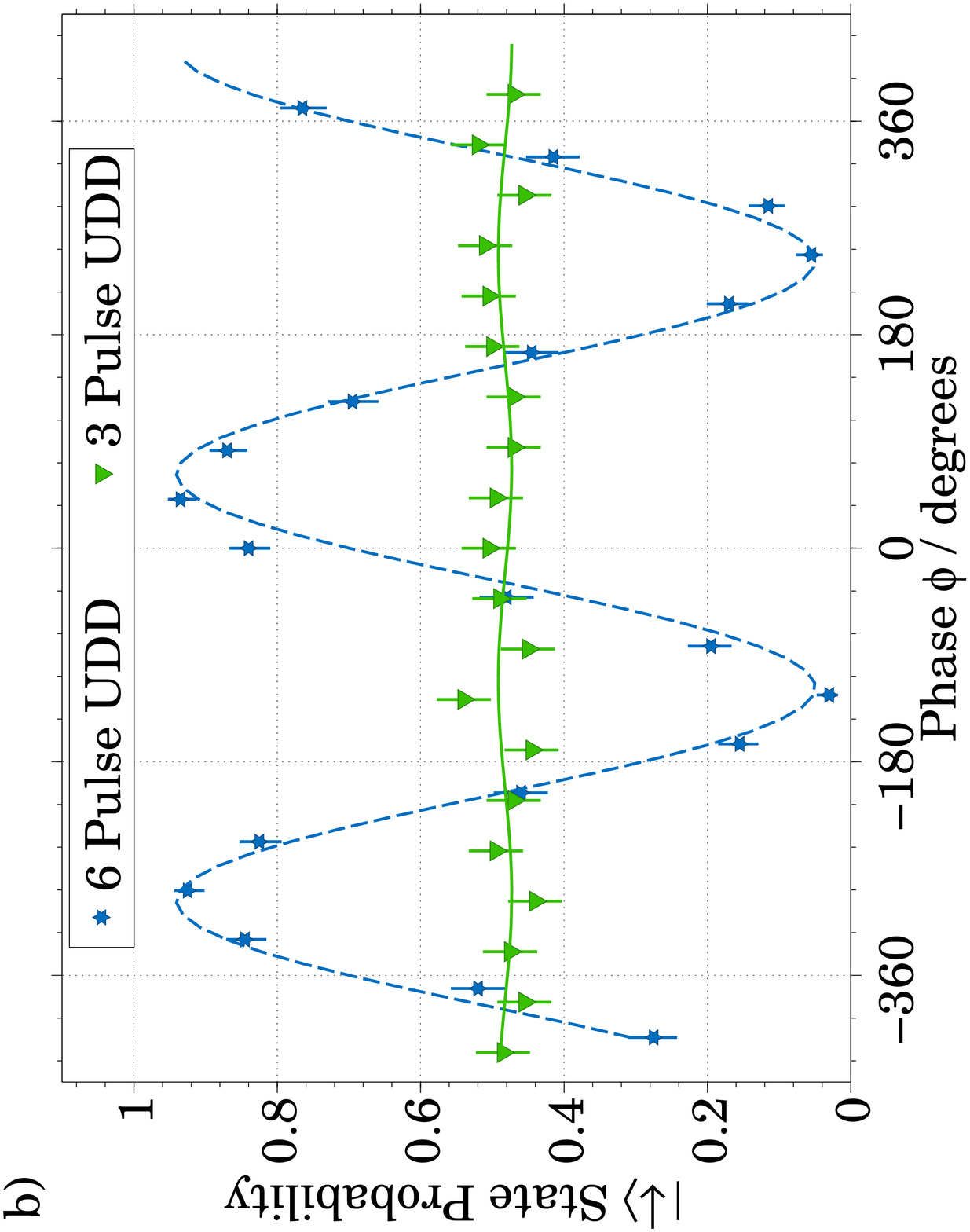}}
	\caption{a) Experimental sequence: the qubit coherence after time $\tau$ is measured by a Ramsey experiment in which the phase $\phi$ of the second \piotwopulse\ is scanned relative to that of the first. An $n=4$ pulse UDD dynamic decoupling sequence is shown in the example (the \piotwopulse\ and \pipulse\ durations are exaggerated for clarity). b) Data (with shot noise error bars) and fitted Ramsey fringes for $n=3$ and $n=6$ pulse UDD sequences, both at $\tau=7\unit{ms}$. The dramatic improvement in the qubit coherence given by the 6-pulse sequence is clear.
\label{fig:EchoSines}}
\end{figure}

Each experiment (Figure \ref{fig:EchoSines}a) starts with the ion optically pumped into state \ketdn.  A decoupling sequence is tested by sandwiching it between two \piotwopulses.  The second pulse has a phase offset $\phi$ relative to the first; scanning this phase leads to Ramsey fringes.  Any loss of phase coherence in the Ramsey gap leads to fringes of reduced contrast, so generally the contrast falls as the gap is made longer.  We aim to show that this fall becomes slower when dynamic decoupling is used.  Finally the qubit state is measured by electron shelving and fluorescence detection, with accuracy up to 99.8\% \cite{Myerson2008}.

The sequence is repeated 200 times for each value of $\phi$, which is typically scanned from $-450\degree$ to $+450\degree$ in 20 steps resulting in the measured state varying sinusoidally with $\phi$. A sine curve is fitted to the data to measure the contrast; example data is shown in figure \ref{fig:EchoSines}b.  Typically 10--20 such runs are taken for a given decoupling sequence, with $\tau$ chosen to be different for each, and with the decoupling pulse timings being scaled accordingly.  

\begin{figure}[t]
	\centerline{\includegraphics[width=0.35\textwidth, angle=-90]{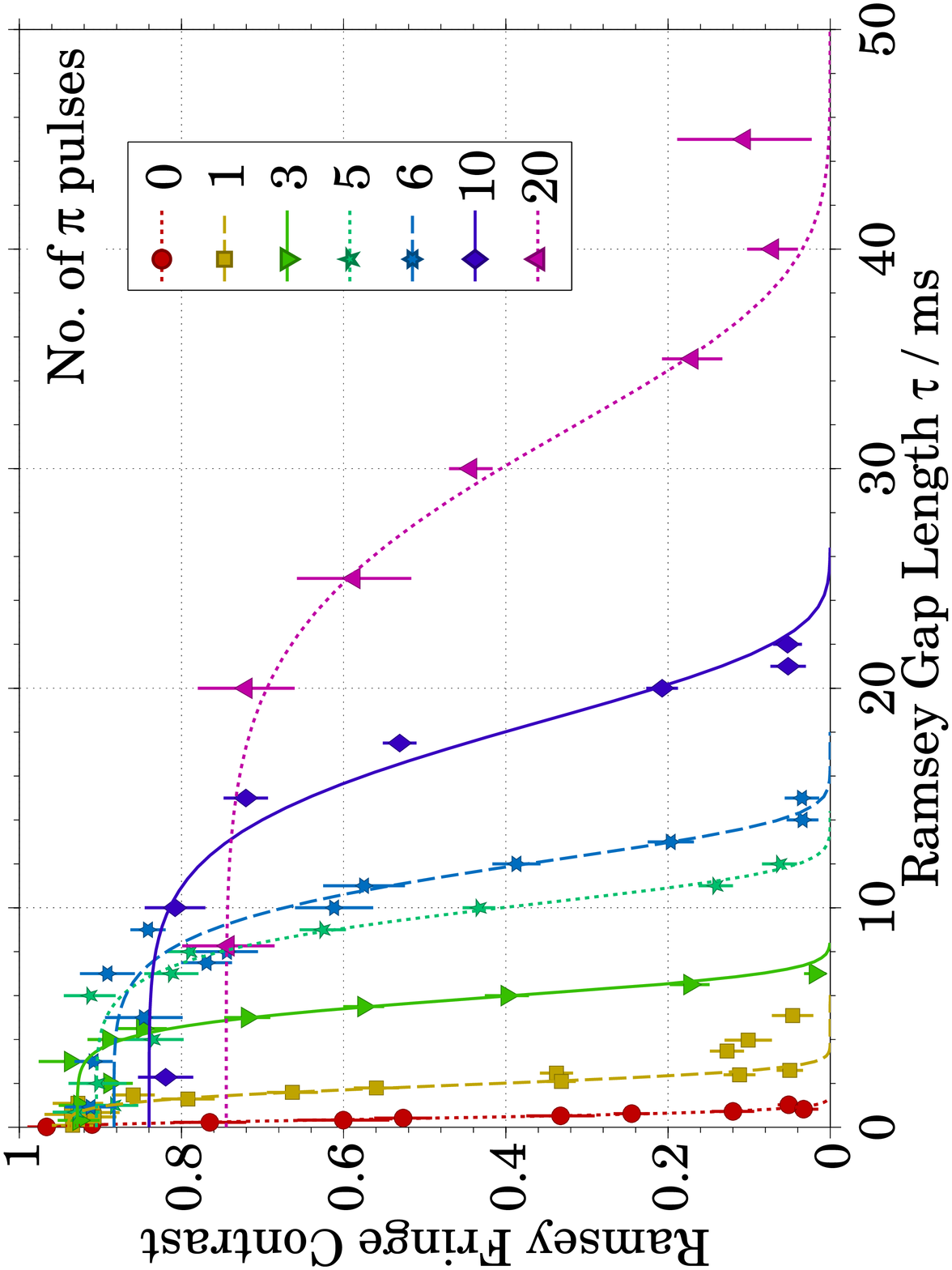}}
	\caption{Comparing UDD sequences with from 0 to 20 \pipulses.  The error bars indicate the uncertainty in each fringe contrast fit (estimated by a bootstrap algorithm).  The curves show the theoretically predicted contrast for each sequence, using \eqn{Contrast} and the fitted $S(\omega)$. Multiple data points with the same $\tau$ have been combined for clarity.
\label{fig:EchoExpt20090420}}
\end{figure}

\begin{figure}[t]
	\centerline{\includegraphics[width=0.35\textwidth, angle=-90]{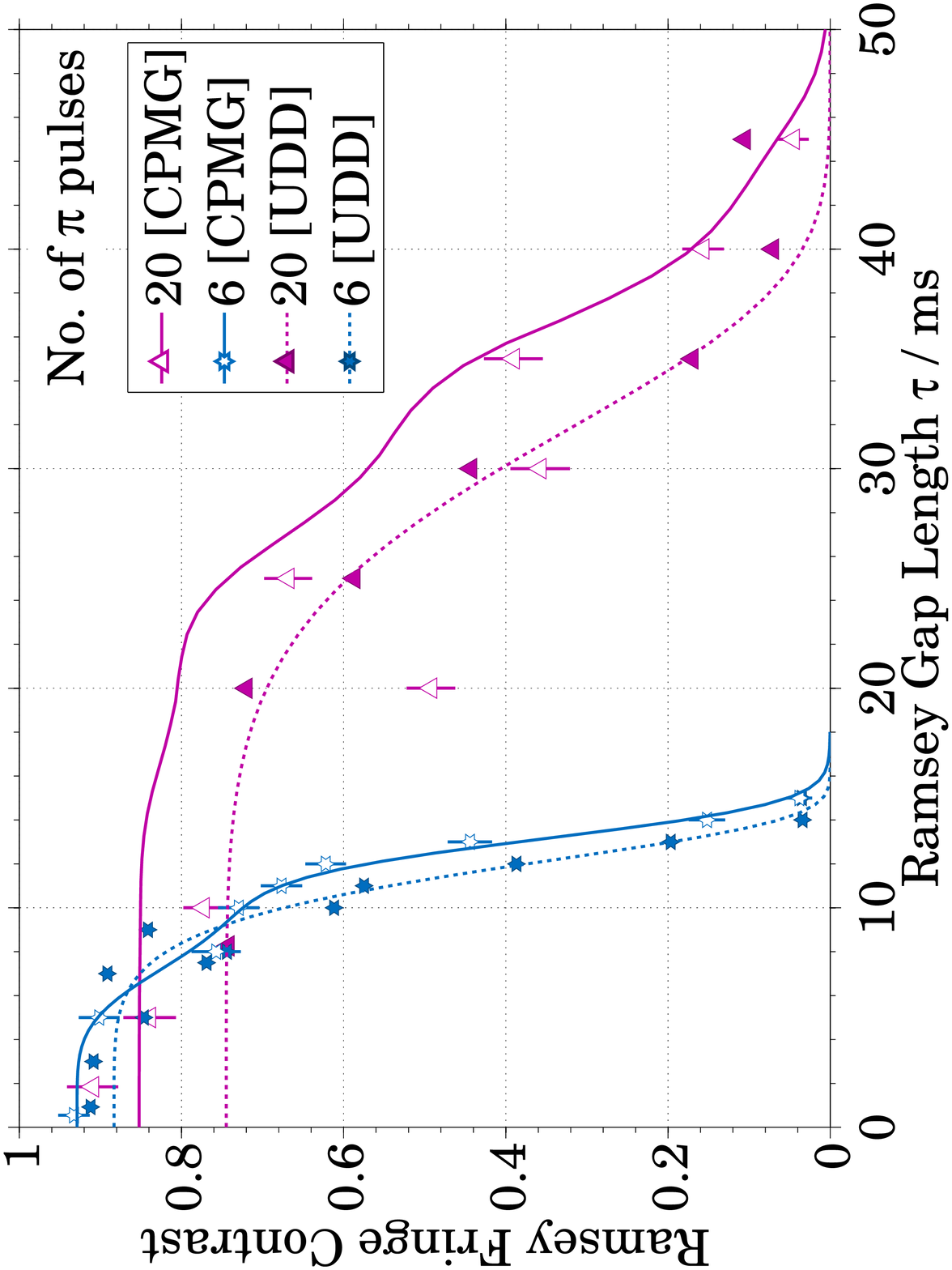}}
	\caption{Comparing UDD and CPMG sequences for six and twenty \pipulses.  Solid symbols (error bars omitted for clarity) and dotted lines are the results and fits from UDD sequences as shown in Figure \ref{fig:EchoExpt20090420}.  Hollow symbols represent CPMG sequences, with solid lines the theoretical prediction using the same fitted noise spectrum $S(\omega)$ as for the UDD sequences.  Data points with the same $\tau$ have been combined for clarity. \label{fig:EchoExpt2009031213}}
\end{figure}

Figure \ref{fig:EchoExpt20090420} shows the results for different numbers of \pipulses. With no dynamic decoupling \pipulses, the fringe contrast drops to $1/\rme$ of its initial value in a time $\tau_c=0.51(5)\unit{ms}$; with a 20-pulse UDD sequence, this time is extended to $\tau_c = 33(1)\unit{ms}$.  We also compared UDD and CPMG (equally spaced \pipulses, at times $\alpha_i = (i-1/2)/n$ for $i=1\ldots n$) sequences, with results shown in figure~\ref{fig:EchoExpt2009031213}. It can be seen that, in our noise environment, UDD performs no better than CPMG; indeed, CPMG is slightly better, extending the coherence time to $\tau_c=37(1)\unit{ms}$ for a 20-pulse sequence, an increase over the unprotected qubit by a factor $\approx 73$, or 1.9 orders of magnitude. The similar performance of UDD and CPMG is expected if the noise spectrum extends to high frequencies; UDD would be superior if the noise spectrum had a sharp high-frequency cutoff \cite{Biercuk2009, Pasini2010}. We also performed experiments both with and without a $90\degree$ phase shift on the UDD \pipulses\footnote{i.e. we tested both CP and CPMG sequences \cite{Carr1954, Meiboom1958}.}, which is equivalent to testing the dynamic decoupling for two different qubit states on the equator of the Bloch sphere; there was no significant difference between the results. 

To fit the data in figures~\ref{fig:EchoExpt20090420} and~\ref{fig:EchoExpt2009031213}, we perform a simulation based on the filter function formalism of Cywi\'nski \myetal\ \cite{Cywinski2008}.  Suppose that the noise power spectrum is given by $S(\omega)$.  We multiply the noise spectrum by the pulse-sequence's filter function $F(\omega t)$ ($F$ also depends on the number and finite duration of the \pipulses\ \cite{Biercuk2009}), and calculate the integral over angular frequency $\omega$:
\begin{equation}
	\chi(t) = \int^\infty_0 \frac{S(\omega)F(\omega t)}{\pi\omega^2} \rmd\omega
\label{eqn:X(t)}
\end{equation}
The qubit coherence $C(t)$ is then given by \cite{Cywinski2008}
\begin{equation}
	C(t) = N\rme^{-\chi(t)}
\label{eqn:Contrast}
\end{equation}
where $N$ is a normalization constant that accounts for effects such as imperfections in the \pipulses\ themselves. In our experiment, $C(t)$ is the contrast of the Ramsey fringes. 

The noise spectrum of the magnetic field measured outside the ion trap vacuum system did not give a good fit to the data when used to calculate $C(t)$, presumably because it differs too greatly from the noise at the position of the ion.  However, we can reverse the process; the dynamically-decoupled ion acts as a spectrometer to measure the field fluctuations \cite{Du2009, Hall2010}.  We model the noise spectrum $S(\omega)$ by a piecewise cubic spline in log-log space, use it to calculate $C(t)$, and find the spectrum which gives the best fit to the experimental data; the fit attempts to match all our UDD and CPMG data with the same $S(\omega)$ (though each data set is allowed its own fitted normalization constant $N$). The calculated contrast $C(t)$ is not very sensitive to the detailed shape of $S(\omega)$, but the procedure does yield a noise spectrum which is close to a power law for $100\Hz \lesssim (\omega/2\pi) \lesssim 100\kHz$, with $S(\omega)\propto \omega^{-5\pm 1}$. This is consistent with the $S(\omega)\propto\omega^{-4}$ spectrum measured by Biercuk \myetal\ inside a superconducting solenoid \cite{Biercuk2008}. The curves in figures~\ref{fig:EchoExpt20090420} and~\ref{fig:EchoExpt2009031213} show the calculated $C(t)$ using this noise spectrum, and fit the experimental data reasonably well.

\begin{figure}[t]
	\centerline{\includegraphics[width=0.35\textwidth, angle=-90]{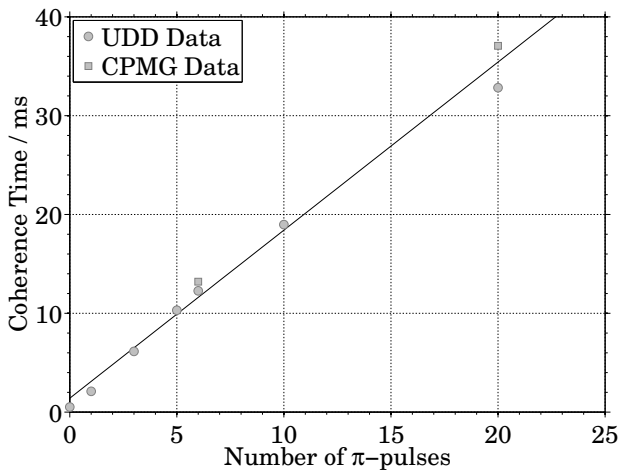}}
	\caption{The $1/\rme$ coherence time, as measured by \eqn{Contrast}, is plotted against the number of \pipulses\ for UDD and CPMG data. A straight line fit to all the data is shown. When the Ramsey delay $\tau$ is equal to the coherence time, the finite-length \pipulses\ occupy a total duration $<0.02\tau$ in all cases.
\label{fig:EchoExpt20090420Times}}
\end{figure}

The fitted $1/\rme$ coherence times are shown in figure~\ref{fig:EchoExpt20090420Times}.  The data is matched well by a straight line, similar to the observations of Ryan, Hodges and Cory \cite{Ryan2010}.

It is clear from figure~\ref{fig:EchoExpt20090420} that although the UDD sequence significantly extends the coherence time of the qubit, the coherence at short time is actually degraded due to imperfections in the \pipulses\ which are more significant the larger the number of pulses used. We estimate the typical \pipulse\ fidelity (based on the fits extrapolated to $\tau=0$) to be 98.7\%. This fidelity could be improved significantly by increasing the Rabi frequency so that it is well above the amplitude $\delta(t)$ of the dominant noise sources, for example by driving the qubit transition with near-field microwaves from electrodes much closer to the ion, as proposed in~\cite{Ospelkaus2008}.

In conclusion, we have shown that extending the Hahn spin-echo to correct for frequency offsets which vary polynomially in time yields the Uhrig dynamic decoupling sequence, and that applying this sequence (or the CPMG sequence) to a single physical qubit stored in a trapped \Ca{43} ion increases the coherence time by nearly two orders of magnitude, to $\tau_c\approx 35\unit{ms}$. In order to demonstrate the increase in coherence time, we chose qubit states in the \lev{S}{1/2} manifold which had the greatest sensitivity to magnetic field fluctuations. For a qubit stored in the magnetic field-insensitive ``clock'' states (4\hfslev{S}{1/2}{3,0} and 4\hfslev{S}{1/2}{4,0}) we have previously measured a coherence time $T_2=1.2(2)\s$~\cite{Lucas2007}; since this was also limited by magnetic field noise, it should be possible to extend the coherence time of such a qubit to several minutes using dynamic decoupling techniques, at which point it becomes practically difficult to measure using a single qubit. The memory qubit coherence time would then exceed the typical timescale for trapped-ion quantum logic gates ($\sim 20\us$ \cite{Kirchmair2009a}) by many orders of magnitude, an essential prerequisite for implementing fault-tolerant quantum computation.

% Acknowledgments
We thank M.~J.~Curtis and D.~N.~Stacey for helpful comments on the manuscript, other members of the Oxford ion trap group for useful discussions, and C.~A.~Ryan for drawing our attention to additional references.  This work was funded by EPSRC (QIP IRC), IARPA (ref.\ 47968-PH-QC), the European Commission (SCALA network) and the Royal Society.

\bibliographystyle{apsrev}
\bibliography{DJS}
\end{document}